\documentstyle[12pt]{article}

\input{psfig}
\begin{document}
\pagenumbering{arabic}
\begin{titlepage}
\vspace*{-6ex}
\begin{flushright}
\fbox{RU 95/E-26}  May 15, 1995
\end{flushright}
\begin{center}
{\large\bf THE STRUCTURE OF THE POMERON} \\
\vspace{2.5cm}
{\large Konstantin Goulianos} \\
\vspace{.5cm}
{\sl The Rockefeller University\\New York, NY 10021, USA}\\
\vspace*{2in}
\begin{abstract}
Results of experiments probing the structure of the
pomeron at hadron colliders and at HERA are reviewed.
By renormalizing the
pomeron flux factor in diffraction dissociation as dictated by unitarity,
a picture emerges from the data in which the pomeron appears to be
made of valence quark and gluon color-singlets in a combination
suggested by asymptopia.
\end{abstract}

\end{center}
\end{titlepage}

\section{Introduction}
The success of Regge theory in describing the main features of high energy
hadronic cross sections with a
{\em universal} pomeron trajectory \cite{KG1,DL1}
has generated considerable interest
in the nature of the pomeron and its QCD structutre.
Since the pomeron has the quantum numbers of the vacuum, it must be
represented by a color-singlet $q\bar q$  and/or gluon
combination of partons.
The question whether or not this combination has a unique hadron-like structure
can ultimately be answered only by experiment.
In this paper we review briefly the results obtained so far by
experiments probing the pomeron constituents
and draw a coherent
conclusion about the partonic structure of the pomeron.

The pomeron structure has been under study in {\em hard} single
diffraction dissociation in hadron colliders and in deep inelastic diffractive
scattering at HERA. Events with a rapidity gap between jets observed
by CDF and D0 are undoubtedly
also related to the pomeron.  The
phenomenology associated with extracting information on the pomeron structure
from these studies relies on Regge theory and factorization.
It is therefore useful to review briefly this phenomenology, particularly
since unitarity requirements that must be imposed on the theory have
a profound effect on the conclusions
that can be drawn from the data about the pomeron structure.

\begin{figure}[htb]
\psfig{figure=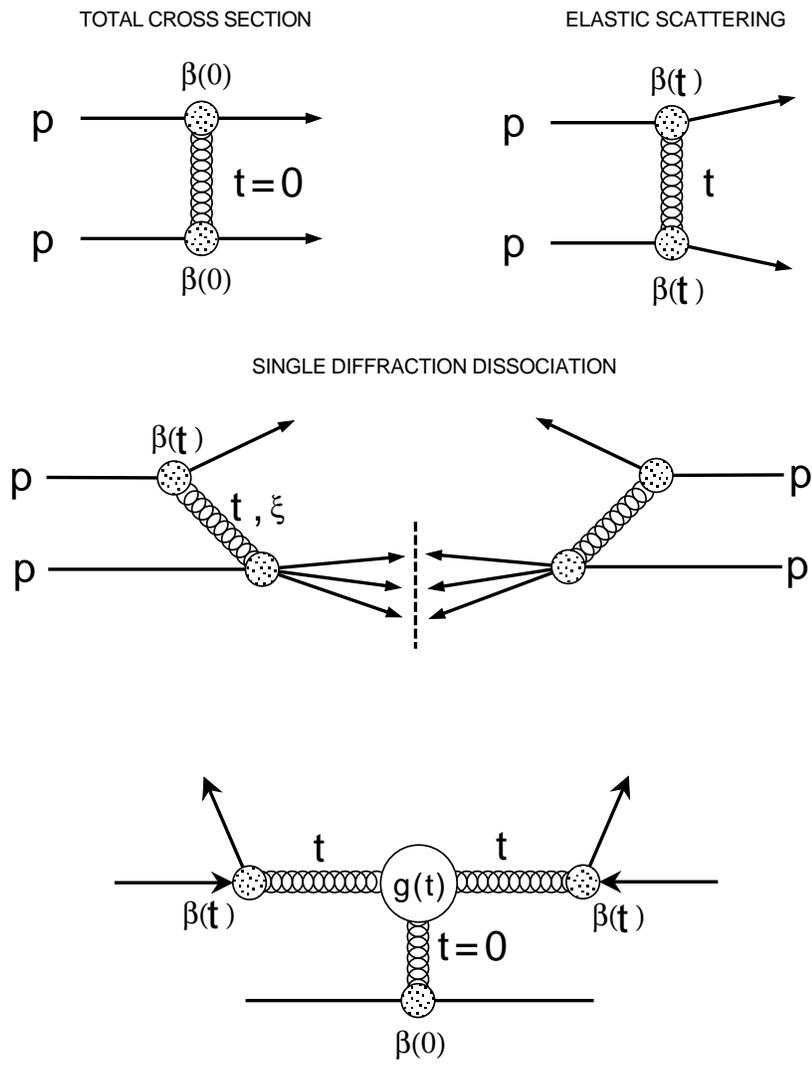,height=8in}
\caption{Feynman diagrams for the total, elastic, and single diffraction
dissociation cross sections, including the ``triple-pomeron" diagram for single
diffraction.}
\label{diagrams}
\end{figure}

The cross section for
single diffraction dissociation
in Regge theory has the form
\begin{equation}
\frac{d^2\sigma_{sd}^{ij}(s,\xi,t)}{dtd\xi}=
\frac{1}{16\pi}\;\beta_{i{\cal{P}}}^2(t)
\;\xi^{1-2\alpha(t)}\;\left[ \beta_{j{\cal{P}}}(0)\,g(t)
\;\left(\frac{s'}{s'_0}\right)^{\alpha(0)-1}\right]
\label{diffractive}
\end{equation}
where ${\cal{P}}$ stands for pomeron, $s'$ is the s-value in the
${\cal{P}}-j$ reference frame,
$s'_0$ is a constant conventionally set to 1 GeV$^2$,
$\xi=s'/s$ is the
Feynman-$x$ of the pomeron in hadron-$i$, and $\alpha(t)$ the pomeron
trajectory given by
\begin{equation}
\alpha(t)=\alpha(0)+\alpha't=1+\epsilon+\alpha't
\label{pomeron}
\end{equation}
Fig.~\ref{diagrams} shows the Feynman diagrams for the total, elastic, and
single
diffractive cross sections, including the ``triple-pomeron" diagram for
single diffraction which is used to derive Eq.~(\ref{diffractive}).
The term in the square brackets in (\ref{diffractive})
may be interpreted as the total cross section of the pomeron with hadron-$j$,
\begin{equation}
\sigma_t^{{\cal{P}}j}(s',t)=\beta_{j{\cal{P}}}(0)\,g(t)
\;\left(\frac{s'}{s'_0}\right)^{\alpha(0)-1}
\label{Pj}
\end{equation}
where $g(t)$ is the pomeron-pomeron coupling,
commonly referred to as the
{\em triple-pomeron} coupling
constant.
Such an interpretation assigns to the pomeron a hadron-like
{\em virtual reality}, which leads naturally to viewing single diffraction
as being due to a flux of
pomerons emitted by hadron-$i$ and interacting with hadron-$j$.
The ``pomeron flux factor", which in this picture depends on
$\beta_{i{\cal{P}}}^2(t)$ and therefore can be obtained
from the $i-i$ elastic scattering differential cross section, is
identified as
\begin{equation}
f_{{\cal{P}}/i}(\xi,t)=\frac{1}{16\pi}\;\beta_{i{\cal{P}}}^2(t)
\;\xi^{1-2\alpha(t)}
\label{flux}
\end{equation}
The assumption of factorization of the flux factor in {\em hard}
processes is coupled to the idea that the pomeron may have a
partonic structure similar to that of hadrons.  However,
even in the absense of such a hadron-like structure,
pomeron exchange must involve partons and therefore
one should expect to observe hard
processes in single diffraction dissociation. Factorization allows such
processes to be
calculated for any particular experiment from the pomeron flux factor and
an assumed partonic structure for the pomeron.  The question of the
uniqueness of the pomeron structure can then be answered by comparing the
results of different experiments with expectations.

A model based on this view was proposed by Ingelman and Schlein (IS) and was
used to calculate high-$P_T$ jet production in $p\bar p$ single diffraction
dissociation \cite{IS}. This calculation was followed by the
discovery of diffractive dijets by UA8 \cite{UA8}. The shape of the
$\eta$-distribution of the jets in the UA8 experiment
favors a
hard structure function for the pomeron, of the type
$F(\beta)=6\beta(1-\beta)$, where $\beta$
is the momentum fraction of a parton inside the pomeron,
over a soft structure function of the type
$F(\beta)=6(1-\beta)^5$.  However, the dijet rates calculated
for such a structure function using the IS model are
substantially higher than the observed ones.
The ``discrepancy factor" required to multiply the pomeron
hard-quark(gluon) structure function
to predict the measured dijet rates
is $0.46\pm0.08\pm0.24$  ($0.19\pm0.03\pm0.10$) \cite{UA8-EPS}.
One possible explanation for this result
is that the {\em virtual} pomeron does not obey the momentum sum rule
\cite{UA8-EPS,DL2}. A more physical explanation,
in which the pomeron {\em obeys}
the momentum sum rule, is offered by
interpreting the pomeron flux as a probability density for finding a pomeron
inside hadron-$i$ and
{\em renormalizing} it
so that its integral is not allowed to exceed unity \cite{KG2}.
Using a renormalized pomeron flux lowers the predicted rates,
thereby increasing the discepancy factors mentioned above
by a factor of $\sim 4$ and bringing the UA8 results into agreement with
the momentum sum rule.

\begin{figure}[htb]
\vspace*{-1in}
\centerline{\psfig{figure=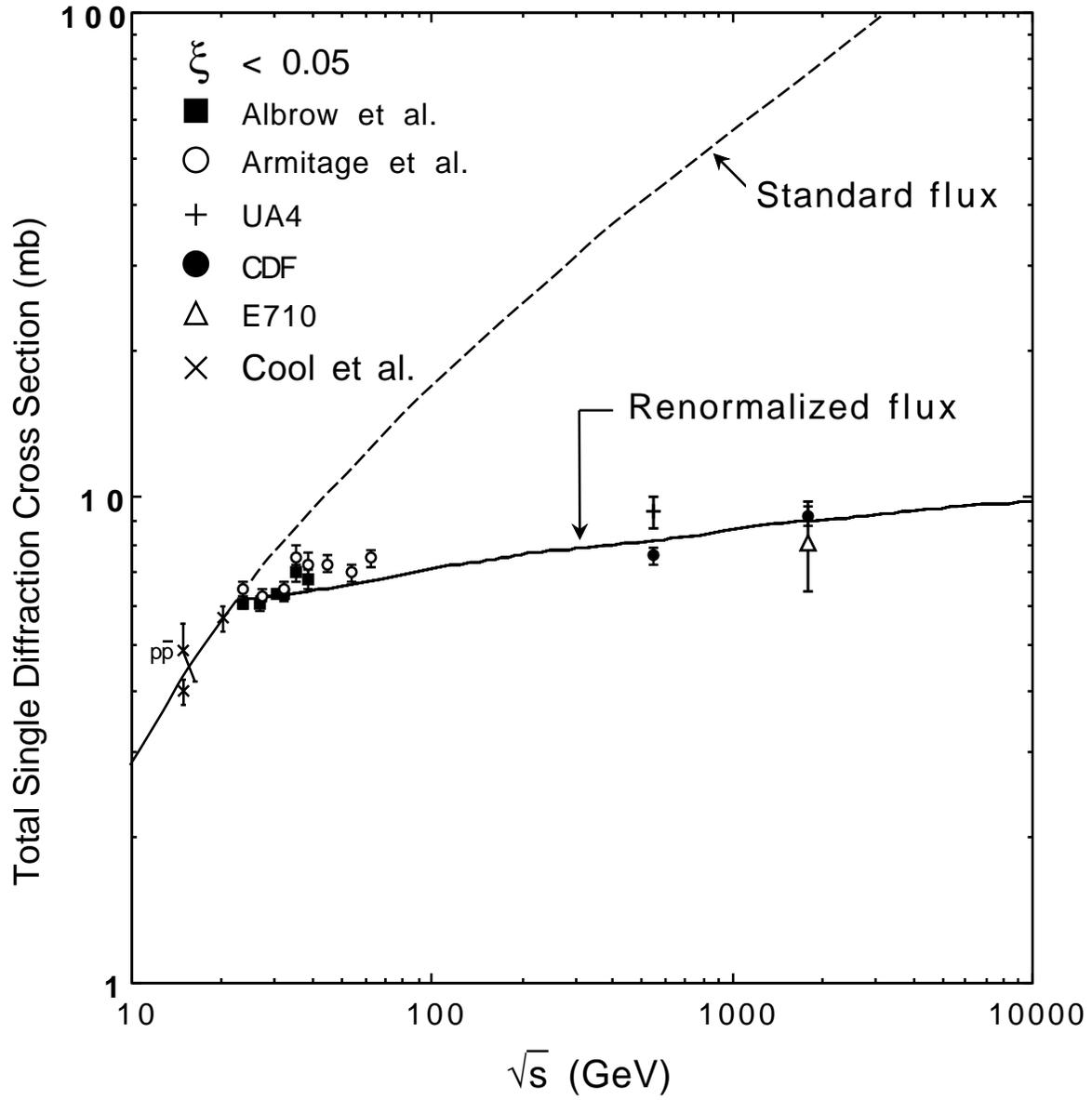,height=10in}}
\vspace*{-1.25in}
\caption{Total \protect{$p(\bar p)-p$}
single diffraction cross section data
for \protect{$\xi<0.05$} compared with the predictions of Regge theory
with a standard and a renormalized pomeron flux.}
\label{sd}
\end{figure}

The  renormalization of the pomeron flux
was  proposed in order
to unitarize the triple-pomeron amplitude, which gives the single
diffractive cross section.  Without unitarization,
the $p\bar p\,$ SD cross section (Eq.~\ref{diffractive})
rises much faster than that observed,
reaching the total cross section and therefore violating unitarity
at the TeV energy scale.
This is shown in Fig.~\ref{sd}, taken from \cite{KG2},
which compares data with
predicted $p\bar p\,$ SD cross sections obtained
with and without a renormalized pomeron flux.
The renormalized flux is given by
$$f_N(\xi,t)=f_{{\cal{P}}/i}(\xi,t)d\xi dt\hspace*{0.5 in}
\mbox{for}\;\;N(\xi_{min})\leq 1$$
\begin{equation}
f_N(\xi,t)=\frac{f_{{\cal{P}}/i}(\xi,t)d\xi dt}{N(\xi_{min})}
\hspace*{0.5 in}\mbox{for}\;\;N(\xi_{min})> 1
\label{FN}
\end{equation}
$$\mbox{with}\hspace*{0.5 cm} N(\xi_{min})\equiv
{\int_{\xi_{min}}^{0.1}d\xi \int_{t=0}^{\infty} f_{{\cal{P}}/i}(\xi,t)dt}$$
where $\xi_{min}$=$(1.5\; \mbox{GeV}^2/s)$ for $p\bar p$ soft single
diffraction.
Below, experimental results on hard diffraction
will be compared with predictions obtained both with the standard and a
renormalized pomeron flux.
\section{Hard diffraction at hadron colliders}
\begin{figure}[htb]
\vspace*{-1in}
\centerline{\psfig{figure=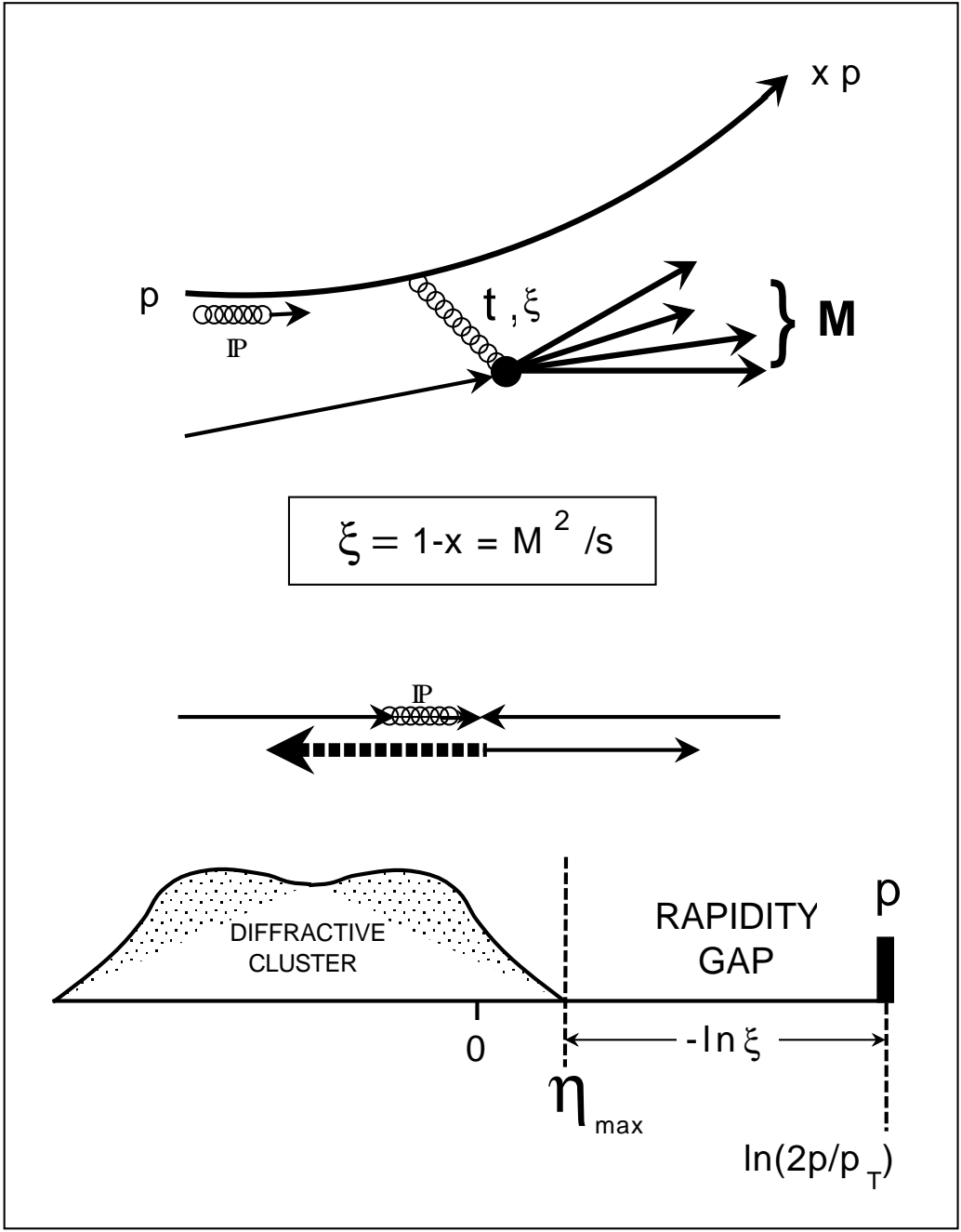,height=8in}}
\vspace*{0.5in}
\caption{Kinematics for $pp$ single diffraction dissociation illustrating the
leading particle and rapidity gap techniques for tagging diffractive events.}
\label{kinematics}
\end{figure}

Single diffraction dissociation provides the most transparent and accessible
window for looking at the structure of the pomeron.  Events are tagged
as diffractive either by the detection of a high-$x_F$
(anti)proton, which presumably ``emitted" a small-$\xi$ pomeron,
or by the presense of a rapidity gap at one end of the kinematic region,
as shown in Fig.~\ref{kinematics}.  Another process involving the pomeron is
hard {\em double}
diffraction dissociation, which is characterized by a rapidity gap
in the central region and one or more jets on each side of the gap.
Below, we review briefly the hard diffraction collider
experiments and discuss the interpretation of their results in terms of a
pomeron structure function.

\subsection{The UA8 experiment}
UA8 pioneered hard diffraction studies
by observing high-$P_T$ jet production in the
process $p+\bar{p}\rightarrow p+Jet_1+Jet_2+X$ at the CERN $Sp\bar{p}S$
collider at $\sqrt{s}=630$ GeV.  Events with two jets of
$P_T>8$ GeV were detected
in coincidence with a high-$x_F$ proton,  whose momentum and angle
were measured in a forward ``roman pot" spectrometer. The event sample spanned
the kinematic range
$$0.9<x_p<0.94\hspace*{1in} 0.9<|t|<2.3\;\mbox{GeV}^2$$
Assuming the jets to be due to collisions between the proton and pomeron
constituents, and comparing the $x_F$ distribution of the sum of the jet
momenta of the events with Monte Carlo
distributions generated with a standard proton
but different pomeron structure functions, UA8 concluded \cite{UA8}
that the partonic
structure of the pomeron is $\sim 57$\% {\em hard} [$6\beta (1-\beta)$],
$\sim 30$\% {\em superhard} [$\delta (\beta)$],
and $\sim 13$\% {\em soft} [$6(1-\beta)^5$].
However,  the dijet
production rate measured by UA8 \cite{UA8-EPS}
is smaller by a factor of $\sim$2(or 5) than the rate
predicted for a pomeron made of hard-quark(or gluon) constituents obeying the
momentum sum rule. As discussed in the introduction, this discrepancy between
the results obtained by the event shape and event rate analyses
was expressed by UA8 in terms of a coefficient by which the full quark or gluon
hard structure function has to be multiplied to yield the
measured rates.  This coefficient, named ``the discrepancy factor", represents
the fraction of the pomeron momentum carried by its partons.  As already
discussed,
with the standard flux normalization the UA8 hard pomeron does not obey the
momentum sum rule.  Using the procedure of pomeron
flux renormalization, the
discrepancy factors of  $0.46\pm0.08\pm0.24$  ($0.19\pm0.03\pm0.10$)
measured by UA8 for
a hard-quark(gluon) dominated pomeron become
$1.79\pm 0.31\pm 0.93\;(0.74\pm 0.11\pm 0.39)$  \cite{KG2}.
These values are both
consistent with unity, so that the momentum sum rule is
restored.  Assuming the momentum sum rule to be exact, the rate analysis
could in principle be used to measure the ratio of the quark to gluon component
of the pomeron. However, the present UA8 results are not accurate enough to
address this issue.

\subsection{Diffractive W's in CDF}
The quark content of the pomeron
can be probed directly with diffractive $W$ production, which to leading order
occurs through $q\bar{q}\rightarrow W$.  A hard gluonic pomeron can also
lead to diffractive $W's$ through $gq\rightarrow Wq (\rightarrow W+Jet)$,
but the rate for this
subprocess is down by a factor of order $\alpha_s$. The ratio of diffractive
to non-diffractive $W^{\pm}(\rightarrow l^{\pm}\nu )$ production has been
calculated by Bruni and Ingelman (BI) \cite{BI}
to be $\sim 17$\% ($\sim 1$)\% for a
hard-quark(gluon), and $\sim 0.4$\% for a soft-quark pomeron structure.
Thus, diffractive $W$ production is mainly sensitive to the hard-quark
component of the pomeron structure function. However, the rates calculated
by BI may be too optimistic.  Using the renormalized pomeron flux
lowers the hard-quark prediction down to 2.8\%  \cite{KG2}.

A search for diffractive $W's$ is currently being conducted by the CDF
collaboration at the Tevatron at $\sqrt{s}=$1800 GeV using
the rapidity gap technique to tag diffraction.
Preliminary results from a study of a sample of $\sim 3,500$ $W$ events
show no signal for diffractive $W$ production at the level of
{\em a few \%} \cite{KG3}, which is to be
compared with the 17\% of the BI and the 2.8\% of the renormalized flux
predictions. This result, therefore,
restricts severely the hard-quark
structure function of the pomeron for the BI-type flux, but
lacks the sensitivity needed to probe the pomeron structure
if the renormalized flux factor is used.

\subsection{Diffractive dijets in CDF}
The rapidity gap method was also used in CDF to search for diffractive
dijet production, which, as in the UA8 experiment, is sensitive to both
the quark and the gluon content of the pomeron.
Because of the higher energy used at the Tevatron, $\sqrt{s}=1800$ GeV
as compared to 630 GeV at the $Sp\bar{p}S$,
dijets in the same diffractive mass-region as UA8,
$M_X^2\sim 150$ GeV$^2$, are produced with
lower pomeron $\xi$, since $\xi \approx
M_X^2/s$.  The signature for such events is two high-$P_T$ jets
on the same side of the rapidity region and a rapidity gap on the other
side.  Since the rapidity gap method integrates over $t$, and because of the
exponential $t$-behavior of the diffractive cross section, the average
$t$-value of the events in CDF is close to zero, in contrast to UA8 for
which $|t|\sim 1.5$ GeV$^2$.  Probing the structure of the pomeron with
the same hard process but different pomeron $\xi$ and $t$ can address
the question of the uniqueness of the pomeron structure.

{}From a study of 3415 events with two jets of $P_T>20$ GeV and
$|\eta|>1.8$, CDF has obtained the preliminary result of $R\leq \;\approx 1$\%
at 95\% CL
for the ratio of diffractive to non-diffractive dijets \cite{AB}, to
be compared with the predictions of $\sim 5$\% and $\sim 0.6$\%
obtained with the
standard and renormalized pomeron
flux for a hard-gluon pomeron structure. Again, this result restricts
the hard partonic component of the pomeron if the standard flux is used
to predict rates,
but places no restrictions if the
renormalized flux is used.
\subsection{Hard double diffraction in CDF and D0}
In double diffraction dissociation both the proton and the antiproton
dissociate by exchanging a pomeron.  The process is characterized by two
diffractive clusters of particles with a rapidity gap in-between. The gap
is due to the {\em colorless} QCD nature of the pomeron, as a result of which
the two diffractive clusters are not color-connected and therefore there is no
radiation between them. A {\em hard} pomeron can also
``kick out" jets into each diffractive cluster
and lead to dijet events with a rapidity gap between the jets.  Such events
have been observed by both CDF and D0.  The fraction of rapidity gap dijet
events (more jets can be present in addition to the {\em leading} jets)
to all dijet events with the same kinematics (same $\eta$-region and $P_T$)
was found to be $R_{jets}=(0.85\pm 0.12^{+0.24}_{-0.12})$\% and $(1.4\pm
0.2)$\%
by CDF \cite{CDF-gap} and D0 \cite{D0-gap}, respectively.  An
estimated rate of (1-3)\% was predicted by Bjorken \cite{Bj}
on purely QCD grounds. A quantitative
connection to double diffraction dissociation was made in \cite{KG2}, where it
was pointed out that the measured rate for $R_{jets}$ is the same within error
as the rate of $R_{soft}=1.2$\% expected for {\em soft} double
diffraction dissociation
in which no jets are present.  The  fact that $R_{jets}=R_{soft}$
suggests that {\em the same hard pomeron}
participates both in soft and in hard diffractive processes.

\section{Deep inelastic diffraction at HERA}
At HERA, the quark content of the pomeron is being probed directly
with virtual high-$Q^2$ photons in $e^-p$ deep inelastic scattering at
$\sqrt{s}\sim 300$ GeV (28 GeV electrons on 820 GeV protons).
Both the Zeus and the H1 Collaborations find that in $\sim (5-10)$\%
of the events
there is a large rapidity gap between the proton and the
other particles, indicating that the virtual photon interacted
with a colorless object ``emitted" by the proton, presumably a pomeron.
The general conclusion arrived at
from the study of these events is that
the pomeron structure is mostly hard, but a substantial soft
component is also present.

Recently, the H1 Collaboration  reported a comprehensive
measurement \cite{H1}
of the diffractive structure function $F^D_2(Q^2,\xi,\beta)$
(integrated over $t$, which is not measured),
where $\beta$
is the fraction of the pomeron momentum carried by the quark being probed.
The measurement was performed in the traditional way
used to measure the
structure function of the proton, but it was done on events with a
rapidity gap.
H1 finds that the $\xi$-dependence factorizes out and that it
can be fit for all $Q^2$ and $\beta$ bins with
the form $1/\xi^{1+2\epsilon}$, which is the same as the expression in the
pomeron flux factor, Eq.~\ref{flux}.
Moreover, the fit yields $\epsilon\approx 0.1$, which is in agreement
with the value measured in {\em soft}
collisions. It therefore appears that {\em the same pomeron} is involved in
hard as in soft collisions, a conclusion that we also reached above in
comparing the results of hard double diffraction dissociation with soft
double diffraction.

In order to obtain a ``picture" of the $\beta$-structure of the pomeron
and its possible
$Q^2$-dependence, H1 integrates the diffractive form factor
$F^D_2(Q^2,\xi,\beta)$ over $\xi$ and provides values for the expression
\begin{equation}
\tilde{F}^D_2(Q^2,\beta)=\int_{0.0003}^{0.05}F^D_2(Q^2,\xi,\beta)d\xi
\label{F-tilde}
\end{equation}
The limits of integration cover the entire range of the experimental
measurements, and the integration was carried out even in the cases where
the lower limit was kinematically inaccessible.
The results for $\tilde{F}^D_2(Q^2,\beta)$ are plotted in Fig.~\ref{HERA}a
as a function of $\beta$ for four $Q^2$-bins: $Q^2=$8.5, 12, 25 and 50 GeV.
Assuming
complete factorization of the flux factor, this figure represents the
pomeron structure function apart from a normalization factor. The structure
appears to be flat in $\beta$ and has a small but
significant $Q^2$ dependence.  However,
these conclusions are altered if one uses the renormalized flux of \cite{KG2}.
As discussed in the introduction, the procedure for flux renormalization
consists in evaluating the integral of the flux factor over the region
$\xi_{min}<\xi<0.1$ and setting it equal to unity if it is
found to be $\geq 1$.  Now, for fixed $Q^2$ and $\beta$, $\xi_{min}=
(Q^2/\beta s)$.  Therefore, the flux integral, which to a good
approximation varies as $\xi_{min}^{-2\epsilon}$, is given by
\begin{equation}
N(s,Q^2,\beta)\approx \left(\frac{\beta s}{Q^2}\;\xi_0\right)^{2\epsilon}=
3.8\left(\frac{\beta}{Q^2}\right)^{0.23}
\label{N}
\end{equation}
where $\xi_0$ is the value of $\xi_{min}$ for which the flux integral is unity.
For our numerical evaluations we use $\sqrt{s}$=300 GeV and a flux factor
with $\epsilon=0.115$ as in \cite{KG2}.
The value of $\xi_0$ turns out to be $\xi_0=0.004$.  Since $\xi_0$
is larger than
$\xi_{min}$ for all points in Fig.~\ref{HERA}a, the flux must be
renormalized for all the points.

The pomeron structure function is obtained from
$\tilde{F}^D_2(Q^2,\beta)$ using factorization:
\begin{equation}
\tilde{F}^D_2(Q^2,\beta)=\left[\frac{\int_{0.0003}^{0.05}d\xi \int_0^{\infty}
dt\;f_{{\cal{P}}/p}(\xi,t)}{N(s,Q^2,\beta)}\right]\,F_2^{{\cal{P}}}(Q^2,\beta)
\label{F2P}
\end{equation}
The expression in the brackets is the normalized flux factor. The integral
in the numerator has the value 2.0 when the flux factor of \cite{KG2} is used.
Eq.~(\ref{F2P}) shows explicitly how factorization breaks down
due to flux renormalization.  The break-down of factorization is a
direct consequence of unitarization.
Assuming now that the pomeron structure function receives contributions
from the four lightest quarks, whose average charge squared is 5/18, the
quark content of the pomeron is given by
\begin{equation}
f^{{\cal{P}}}_{q}(Q^2,\beta)=\frac{18}{5}\,F_2^{{\cal{P}}}(Q^2,\beta)
\label{FQ}
\end{equation}
The values of $f^{{\cal{P}}}_q(Q^2,\beta)$ obtained in this manner are
shown in Fig.~\ref{HERA}b. As seen, the renormalized points show no $Q^2$
dependence. We take this fact as an indication that
the pomeron {\em reigns in the kingdom of asymptopia}
and compare the data points with the asymptotic
momentum fractions expected for any quark-gluon construct
by leading-order perturbative QCD, which for
$n_f$ quark flavors are
\begin{equation}
f_q=\frac{3n_f}{16+3n_f}\hspace*{0.5in} f_g=\frac{16}{16+3n_f}
\label{flavors}
\end{equation}
The quark and gluon components of the pomeron structure
are taken to be
 $f_{q,g}^{{\cal{P}}}(\beta)=f_{q,g}\;[6\beta(1-\beta)]$.
For $n_f=4$, $f_q=3/7$ and $f_g=4/7$. The pomeron in this picture is
a combination of valence quark and gluon color-singlets and its complete
structure function, which obeys the momentum sum rule, is given by
\begin{equation}
f^{{\cal{P}}}(\beta)=\frac{3}{7}[6\beta(1-\beta)]_q+
\frac{4}{7}[6\beta(1-\beta)]_g
\label{fP}
\end{equation}
The data in Fig.~\ref{HERA}b are in
reasonably good
agreement with the quark-fraction of the structure function given by
$f_{q}^{{\cal{P}}}(\beta)=(3/7)[6\beta(1-\beta)]$, except for
a small excess at the low-$\beta$ region.  An excess at low-$\beta$
is expected in this
picture to arise from
interactions of the photon
with the gluonic part of the pomeron through gluon splitting into
$q\bar q$ pairs.  Such interactions, which are expected to be down by an order
of $\alpha_s$, result in an {\em effective} quark $\beta$-distribution
of the form $3(1-\beta)^2$. We therefore compare in Fig.~\ref{HERA}b
the data with the
distribution
\begin{equation}
f_{q,eff}^{{\cal{P}}}(\beta)=(3/7)[6\beta(1-\beta)]+\alpha_s(4/7)[3(1-\beta)^2]
\label{fqeff}
\end{equation}
using $\alpha_s=0.1$. Considering that this distribution involves
{\em no free parameters}, the agreement with the data is remarkable!

\section{Summary and conclusion}
We have reviewed the experimental measurements on hard diffraction
at hadron colliders and on deep inelastic scattering
with large rapidity gaps at HERA,
and have derived from the data a structure function for the
pomeron. Using the Ingelman-Schlein model, in which a flux of pomerons
``emitted" by the nucleon interacts with the other nucleon at hadron
colliders or with a virtual high-$Q^2$ photon at HERA, the picture of the
pomeron structure that
emerges depends on the normalization of the pomeron flux. Two expression
for the flux
were used: the standard flux used in the literature,
and the {\em renormalized}
flux of \cite{KG2}. Renormalizing the pomeron flux was proposed as a means
of unitarizing the triple-pomeron amplitude.
Our conclusions
do not depend crucially on the particular parametrization of the
standard flux, but the process of renormalization alters the picture
drastically.

\begin{figure}[htb]
\vspace*{-1in}
\centerline{\psfig{figure=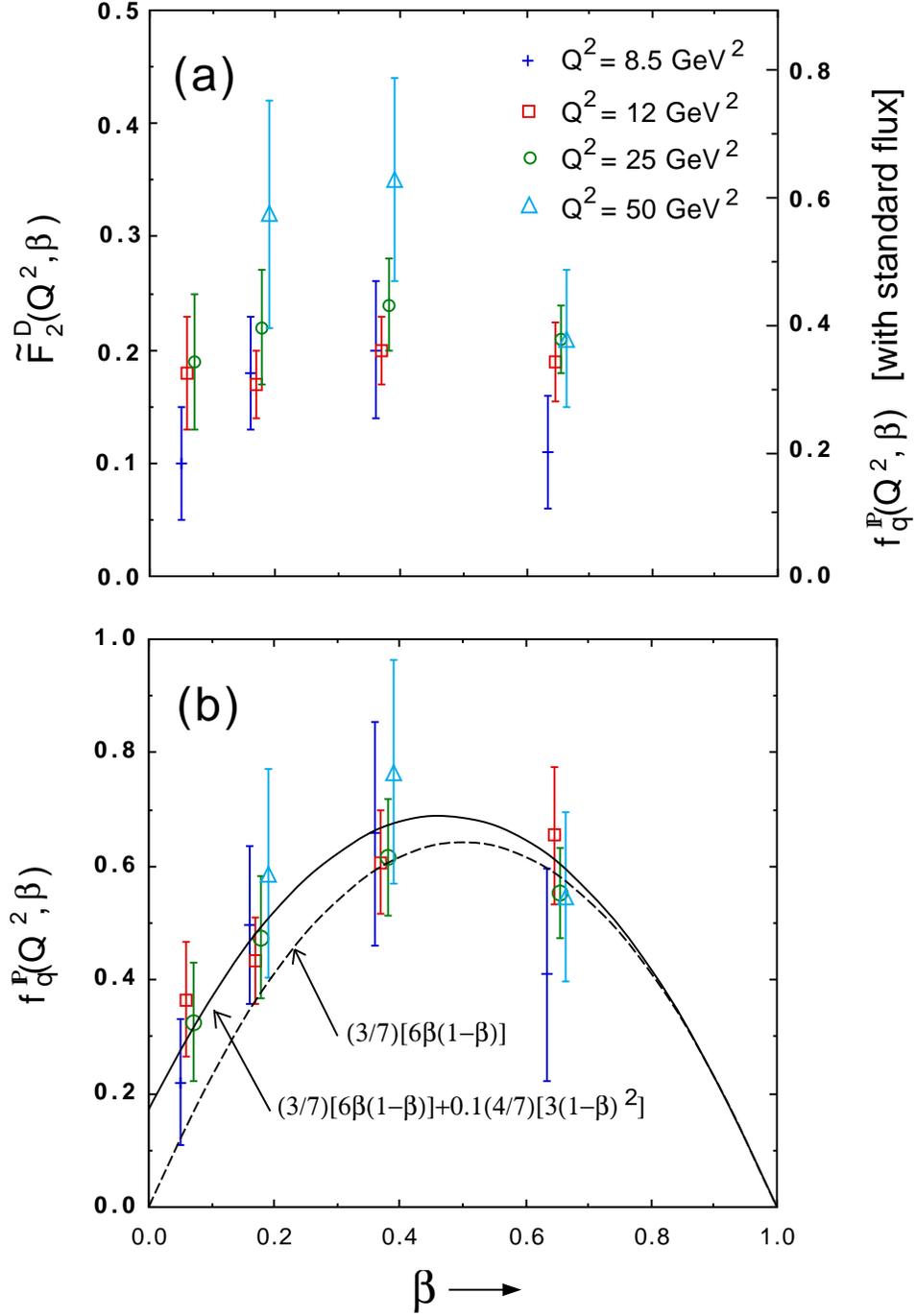,height=10in}}
\vspace*{-1.25in}
\caption{(a) The diffractive structure function measured by H1 at HERA
(see Eq.~6);  the right-hand y-axis gives the pomeron quark content
obtained with the standard flux assuming 4 quark flavors.
(b)  The pomeron quark structure function obtained using the renormalized
pomeron flux.}
\label{HERA}
\end{figure}

With the standard flux, the quark component of the pomeron at HERA is given
by $1.8\,\tilde{F}_2^D(Q^2,\beta)$ (right-hand axis in Fig.~\ref{HERA}a),
where the factor of 1.8 is 18/5
divided by the integral of the standard flux factor, which is 2.0 in our
parametrization. The $\beta$-dependence in Fig.~\ref{HERA}a is rather flat, and
$f_q^{{\cal{P}}}(Q^2,\beta)$
integrates out to an average
value of $\bar{f}_q\sim 1/3$. In contrast, UA8 finds a hard structure
with very little room for a soft component. Also, a 1/3 hard-quark component
would almost saturate the UA8 rate, leaving little room for gluons
in the pomeron.
Coming now to the CDF results, with such a structure one would
predict a diffractive to non-diffractive W fraction
of $\sim 6-8$\%, depending on the
flux parametrization, which is to be compared with the preliminary result of
less than {\em a few \%}.
Thus, the standard flux presents a picture of a mostly
quark-made pomeron, which does not satisfy the momentum sum rule and
is struggling to satisfy the experimenters
of HERA, UA8 and CDF.

Flux renormalization restores order by presenting us with a pomeron that
obeys the momentum sum rule and satisfies all present experimental
constraints.  This pomeron consists of a combination of valence
quark and gluon color-singlets
in a ratio suggested by asymptopia for four quark flavors.
In detail, the results obtained with this model are:
\begin{itemize}
\item No free parameters are needed to fit the HERA data.
\item HERA and UA8 both find a predominantly
hard structure with a small soft component, which can be accounted for by
gluon-splitting into $q\bar q$ pairs
or gluon radiation by the quarks of the pomeron.
\item For a pomeron consisting of 3/7 quark and 4/7 gluon hard components, the
discrepancy factor for UA8 becomes $1.19\pm 0.18\pm 0.61$, which is
consistent with unity and therefore in agreement with the momentum sum rule.
\item The diffractive $W$ production fraction at the Tevatron
is predicted to be 1.2\%.  This value
is not in conflict with the CDF null result of {\em a few \%} accuracy.
\item The diffractive dijet fraction
at the Tevatron for jet $P_T>20$ GeV and $|\eta|>1.8$
is predicted to be 0.5\%, which is also not in conflict
with the CDF measurement.
\end{itemize}
In conclusion, a pomeron structure function as given by Eq.~(\ref{fP})
accounts for
all present experimental results when used in conjunction with the
renormalized flux of \cite{KG2}.

\newpage
\clearpage

\clearpage

\clearpage

\end{document}